\UseRawInputEncoding 

\documentclass[twocolumn,showpacs,floats,floatfix,superscriptaddress,aps,pra]{revtex4-1}
\usepackage{amsfonts}
\usepackage{amssymb}
\usepackage{amsmath}
\usepackage{calc}
\usepackage{graphicx}
\usepackage{bm}

\usepackage[normalem]{ulem}
\usepackage{amsmath,amssymb}
\usepackage{multirow}
\usepackage{xcolor,soul}
\usepackage{srcltx}
\usepackage{hyperref,graphicx}

\def\be{ \begin{equation}}
\def\ee{ \end{equation}}
\def\bea{ \begin{eqnarray}}
\def\eea{ \end{eqnarray}}
\def\bse{ \begin{subequations}}
\def\ese{ \end{subequations}}
\def\bc{ \begin{center}}
\def\ec{ \end{center}}

\newcommand{\stef}[1]{{\color{black} #1}}

\begin{document}

\author{Stefano Longhi$^{*}$} 
\affiliation{Dipartimento di Fisica, Politecnico di Milano, Piazza L. da Vinci 32, I-20133 Milano, Italy}
\affiliation{IFISC (UIB-CSIC), Instituto de Fisica Interdisciplinar y Sistemas Complejos, E-07122 Palma de Mallorca, Spain}
\email{stefano.longhi@polimi.it}

\title{Spectral deformations in non-Hermitian lattices with disorder and skin effect: a solvable model}
  \normalsize


%
\bigskip
\begin{abstract}
\noindent  
 We derive analytical results on energy spectral phase transitions and deformations in the simplest model of one-dimensional lattice displaying the non-Hermitian skin effect, namely the Hatano-Nelson model with unidirectional hopping, under on-site potential uncorrelated disorder in complex energy plane. While the energy spectrum under open boundary conditions (OBC) exactly reproduces the distribution of on-site potential disorder, the energy spectrum under periodic boundary conditions (PBC) undergoes spectral deformations, from one or more closed loops in the fully delocalized phase, with no overlap with the OBC spectrum, to a mixed spectrum (closed loops and some OBC energies) in the mobility edge phase, to a complete collapse toward the OBC spectrum in the bulk localized phase. Such transitions are observed as the strength of disorder is increased.
  Depending on the kind of disorder, different interesting behaviors are found. In particular,  for continuous disorder with a radial distribution in complex energy plane it is shown that in the delocalized phase the energy spectrum under PBC is locked and fully insensitive to disorder, while transition to the bulk localized phase is signaled by the change of a topological winding number. When the disorder is described by a discrete distribution, the bulk localization transition never occurs, while topological phase transitions associated to PBC energy spectral splittings can be observed.
\end{abstract}

\maketitle
      
\section{Introduction}
The interplay between non-Hermiticity, topology and disorder is attracting a great interest in different areas of physics, ranging from condensed matter and mesoscopic physics to cold atoms and classical systems as photonic, acoustic, electric and mechanical systems (for recent reviews see \cite{U1,U2,U3,U4,U5,U6}). Such studies have been largely motivated by the recent experimental advances in engineering dissipation in synthetic matter, offering a great flexibility in molding non-Hermitian Hamiltonians \cite{Em1,E0,E1,E2,E3,E4,E5,E6}.\\
Even in the absence of disorder, non-Hermitian lattice systems show a wide variety of unique and exotic phenomena, which have been investigated in many recent works  \cite{r1,r2,r3,r4,r5,r6,r7,r8,r9,r10,r11,r12,r13,r14,r15,r16,r17,r18,r19,r19b,r20,r21,r22,r23,r24,r25,r26,r27,r28,r29,r30,r31,r32,r33,r34,r35,r36,r37,r38,r39,r40,r41,r42,r42b,r43,r43a,r44,r45,r46,r47,r47b,r47c}. These include the strong sensitivity of the energy spectrum to boundary conditions, the non-Hermitian skin effect, i.e. the extensive number of boundary modes under
the open boundary conditions, the breakdown of the bulk-boundary correspondence based on Bloch band topological invariants, phase transitions of non-Bloch bands, anomalous dynamics and transport under external driving In this context, a wide variety of excitng results have been presented, opening up different research directions. For example, non-Bloch band theory has been introduced to describe the skin effect and to restore the bulk-boundary correspondence \cite{r8,r10,r11,r17,r18,r32}, highlighting the topological origin of the non-Hermitian skin effect \cite{r29,r33}. Different forms of the non-Hermitian skin effect have been predicted, such as the reciprocal skin effect \cite{E4}, the critical skin effect \cite{r38}, the higher-order skin effect \cite{r14,r41,r42,r42b}, and the skin effect in open quantum systems \cite{r16,r43,r44,r45}. Non-Hermiitan skin effect induced by  static or fluctuating disorder has been also predicted \cite{E5,r46,r47}.\\
 The impact of non-Hermiticity in crystalline systems with disorder has been investigated in several works as well (see, for instance, \cite{h1,h2,h3,h4,h5,h6,h7,h8,h9,h10,h11,h14,h14b,h15,h16,h17,h18,h19,h20} and references therein). A landmark model revealing the interplay between the skin effect and disorder is provided by the Anderson model of localization in a one-dimensional lattice with asymmetric hopping amplitudes \cite{U1,h1,h2,h3,h6}, introduced by Hatano and Nelson  in a few pioneering works more than two decades ago \cite{h1,h2}. Hatano and Nelsons predicted the existence of mobility edges and a delocalization transition induced by an imaginary gauge field under uncorrelated random disorder. Recently, the delocalization transition of the Hatano-Nelson model was related to the inherent topology of energy spectra in complex plane \cite{U1,r46,h7,h9,h20}, and extended to systems with aperiodic order \cite{
h7,h8,h9,h15,h16,h17,h18,h19,h20}. 
 However, there are some issues that have been barely studied, for example how the energy spectra of disordered lattices displaying the skin effect are influenced by boundary conditions and how these spectra deform as the disorder strength is increased to approach the bulk localized phase. Unfortunately, most of studies in disordered systems have to rely on large-scale numerical simulations, and there are very few exactly solvable models. A noticeable exception is provided by the Hatano-Nelson model in the limiting case of unidirectional hopping \cite{U1}, which is amenable for some analytical treatment. \stef{ The realization of synthetic lattices with asymmetric hopping and controlled disorder, demonstrated in recent experiments \cite{E5,h21,h21b}, have stimulated a renewed interest in the understanding of the interplay among non-Hermiticity, topology and disorder \cite{h22,h23,h24,h25} with potential impact to applications, such as in the design of non-Hermitian topological classical or quantum sensors \cite{h26}.\\
 In this work we unravel an interesting interplay between disorder, spectral deformations and bulk localization/delocalization transitions in an exactly-solvable model of non-Hermitian system displaying the non-Hermitian skin effect, with some surprising results such as the spectral resilience against certain type of disorder in systems with periodic boundary conditions. Specifically, we}
 study theoretically the Hatano-Nelson model with unidirectional hopping and with on-site potential disorder, providing general results on energy spectra under periodic boundary conditions (PBC) and open boundary conditions (OBC). We unveil some unexpected results, for example strong robustness of PBC energy spectra under a wide class of uncorrelated disorder satisfying rotational invariance in complex plane and 
the existence of topological phase transitions without any bulk localization in any type of disorder with a discrete probability distribution.

\section{Energy spectra of the Hatano-Nelson model with unidirectional hopping}
The simplest non-Hermitian model displaying the non-Hermitian skin effect (NHSE) is provided by the Hatano-Nelson model \cite{U1,h1}, which is described by the equation
\begin{equation}
H \psi_n \equiv J_R \psi_{n-1}+J_L \psi_{n+1}+V_n \psi_n=E \psi_n
\end{equation}
for the wave function amplitude $\psi_n$ at the $n$-th site in the lattice, where $J_{L,R}$ are the left/right hopping amplitudes and $V_n$ is the on-site potential that accounts for diagonal disorder in the system
We assume that $V_n$ are independent complex random variables with the same probability distribution $f(V)$ in complex energy plane $V$ with non-vanishing amplitude inside a finite domain $\Omega$. For a lattice comprising $N$ sites ($n=1,2,3,...,N$), the matrix Hamiltonian $H$ in Eq.(1) is denoted by $H_{OBC}$ when we consider the OBC of a linear chain
\[
\psi_0=\psi_{N+1}=0,
\]
 and by $H_{PBC}$ when we consider the PBC on a ring
\begin{equation}
\psi_{n+N}=\psi_n \exp(i \Phi)
\end{equation}
where $\Phi$ is an external magnetic flux threading the ring (Fig.1). The thermodynamic limit corresponds to the large $N$ limit.\\ 
 In the absence of disorder, i.e. for $V_n=0$, the energy spectrum under PBC reads 
 \[ 
 E(q)=J_R \exp(iq+i \Phi/N)+J_L \exp(-iq-i \Phi/N), \]
  with $q=q_l= 2 \pi l /N$ and $l=1,2,3,...,N$. In the continuous limit, as the quasi-momentum $q$ spans the Brillouin zone, the PBC energy spectrum describes an ellipse in complex energy plane, which collapses into a segment on the real energy axis in the Hermitian limit $J_L=J_R$. For asymmetric hopping $J_R \neq J_L$, the energy spectrum is strongly sensitive to the boundary conditions, and under OBC the spectrum collapses to a segment on the real axis and one observes the NHSE, with all eigenstates squeezed toward the left or right edges of the lattice.  A topological phase transition, separating skin state localization at left or right edges,  occurs at $J_R=J_L$ \cite{U1}. \\ 
The skin effect under OBC persists in the presence of on-site potential disorder, until for sufficiently strong disorder all eigenstates become localized in the bulk and OBC/PBC energy spectra do coincide \cite{U1}. Such a localization mechanism in the bulk induced by disorder, which counteracts the NHSE, is associated to a spectral deformation of the PBC energy spectrum to finally collapse and coincide with the OBC energy spectrum in the fully localized phase. Even for the simple one-band model with nearest-neighbor hopping described by Eq.(1), there are not known exact results for spectral deformations and collapse of PBC and OBC energy spectra under on-site potential disorder, and available results resort to numerical diagonalization of matrices $H_{OBC}$ and $H_{PBC}$ (see for instance, \cite{U1,h1,h2,h3}). Here we focus our attention to the limiting case of the Hatano-Nelson model corresponding to unidirectional hopping on the lattice, i.e. either $J_L=0$ or $J_R=0$, which is amenable for an exact analytical study. This model, albeit being extremely simple and far from catching the entire complexity of other non-Hermitian models, hinders a variety of interesting effects that shed new light on the interplay between skin effect, disorder and localization. In the following, we will assume $J_L=0$ and set $J_R=J$. Note that, in the absence of disorder, in this limiting case the energy spectrum under PBC describes a circumference of radius $J$ in complex energy plane, while the energy spectrum under OBC collapses to the single point of energy $E=0$, which is a high-order exceptional point of the matrix Hamiltonian $H_{OBC}$ with a single eigenvector $\psi^{(0)}=(0,0,0,0,...,1)^T$, fully squeezed toward the right edge of the lattice.
 
\subsection{Energy spectrum under open boundary conditions}
Under OBC, the matrix $H_{OBC}$ associated to Eq.(1) is upper triangular, so that the energy spectrum $E$ is simply given by the diagonal elements $V_n$ of the matrix. Hence, the energy spectrum in the presence of disorder, under OBC, exactly reproduces the distribution of on-site potential disorder. The main effect of disorder is to remove, partly o fully, the degeneracy and defective nature of the zero-energy exceptional point. 
Rather generally, the system under OBC can be found in three different phases: the {\em skin phase}, the {\em mixed phase} and the {\em bulk localized phase}. In the {\em skin phase}, all eigenvectors of $H_{OBC}$ are squeezed toward the right edge of the lattice as in the disorder-free system. This phase usually corresponds to a weak disorder in the system, so as the tendency of  bulk localization induced by the on-site potential disorder is not strong enough to prevent the skin effect. In the {\em mixed phase} some eigenvectors of $H_{OBC}$ are localized in the bulk and some others at the right edge of the lattice, i.e. we observe coexistence of bulk localized modes and skin modes. This phase corresponds usually to a moderate disorder in the system. Finally, in the {\em bulk localized phase} almost all eigenvectors of $H_{OBC}$ are localized 
in the bulk. This phase is usually observed for strong disorder in the system, so as the bulk localization tendency induced by on-site potential disorder overcomes the skin effect.\\ 
Two different cases should be 
distinguished, depending on whether $V$ is a discrete or continuous random variable.\\
(i) {\em Discrete random disorder.} In this case $V$ takes a set of discrete values $W_1$, $W_2$, ..., $W_M$ with probabilities $p_1$, $p_2$, .... $p_M$ ($\sum_l p_l=1$), i.e. $f(V)= \sum_{l=1}^N p_l \delta(V-W_l)$. An important example is provided by a binary (Bernoulli) disorder. The energy spectrum of $H_{OBC}$ is formed by the set of $M$ distinct values $W_1$, $W_2$,..., $W_M$, and the degeneracy is only partly removed, i.e. $H_{OBC}$ remains defective since there are only $M$ linearly independent eigenvectors each associated to the  distinct $M$ eigenenergies. Like in the disorder-free lattice, the spatial distributions of such linearly-independent eigenvectors display the NHSE, i.e. they are squeezed toward the right edge of the lattice. Namely, let $\psi^{(W_l)}$ be the eigenvector associated to the eigenenergy $E=W_l$,  and let $n_l$ be the largest site index in the lattice where the potential $V_n$ at $n=n_l$ takes the value $W_l$. Then, apart from a normalization factor, one has
\stef{
\begin{equation}
\psi_n^{(W_l)}= \left\{
\begin{array}{lll}
0 & {\rm for} & 1 \leq n < n_l \\
1 & {\rm for} & n=n_l \\
\prod_{s=n_l+1}^{n} \frac{J}{W_{l}-W_s} &{\rm for} &  n_l < n \leq N
\end{array}
\right.
\end{equation}
}
 \begin{figure}[htbp]
  \includegraphics[width=82mm]{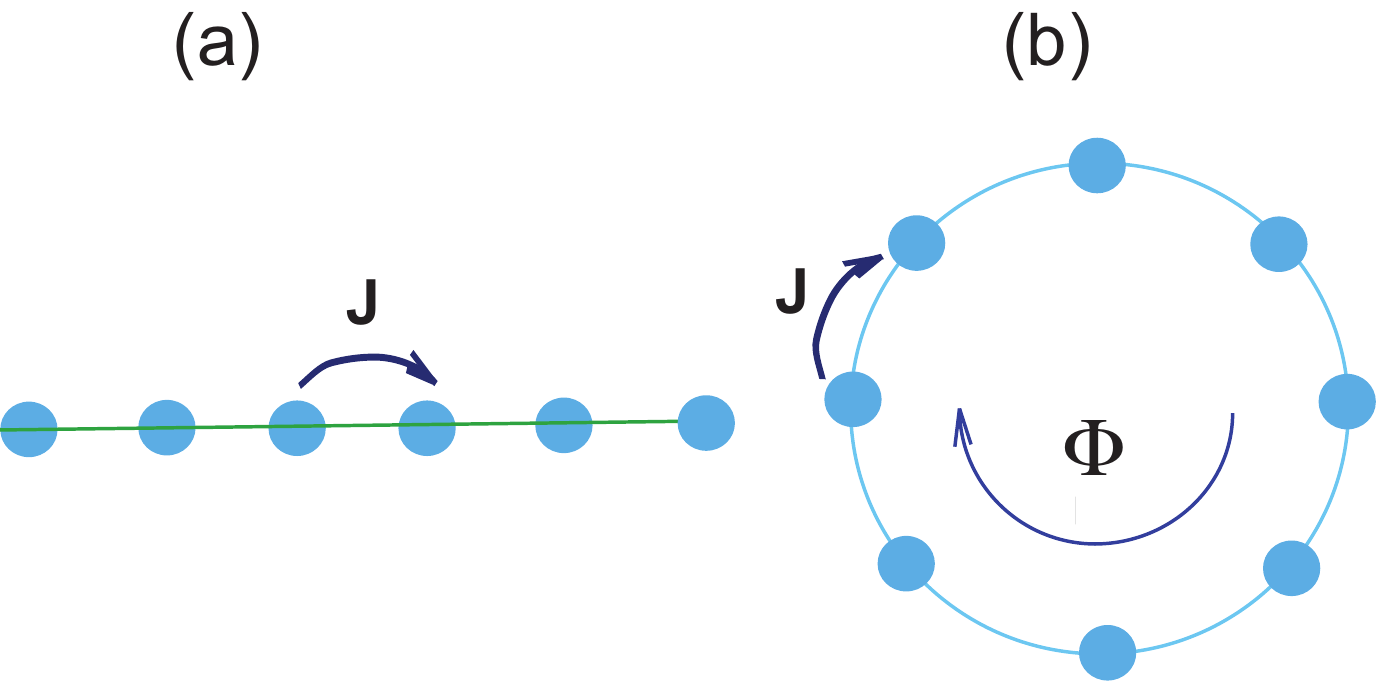}\\
  \caption{(color online) Schematic of the non-Hermitian lattice with unidirectional hopping under (a) OBC (linear chain geometry), and (b) PBC (ring geometry with a magnetic flux $\Phi$).}
\end{figure}
For example, let us consider a lattice comprising $N=18$ sites and the Bernoulli distribution of on-site potential disorder, so that $V$ can take only two values $0$ and $W$ with some probabilities $p_1=p$ and $p_2=1-p$. Consider, as an example, the following realization of disorder 
\begin{equation}
 W,0,W,W,0,W,W,0,0,W,W,W,W,0,W,W,0,0.
\end{equation}
Then there are two non-degenerate eigenvalues $E=0$ and $E=W$ of $H_{OBC}$, with two corresponding linearly-independent eigenvectors given by
\begin{equation}
\psi^{(0)}=(0,0,0,0,0,0,0,0,0,0,0,0,0,0,0,0,0,1)^T
\end{equation}
and 
\stef{
\begin{equation}
\psi^{(W)}=\left(0,0,0,0,0,0,0,0,0,0,0,0,0,0,0,1,\frac{J}{W}, \frac{J^2}{W^2} \right)^T
\end{equation}
}
respectively. Note that, regardless of the strength of disorder, \stef{ in the thermodynamic limit $N \rightarrow \infty$ all the eigenvectors  $\psi_n^{(W_l)}$ are squeezed near the right edge of the lattice, with vanishing occupation amplitudes of all sites from $n=1$ to $n=n_{l_0}$, where $n_{l_0}$ is the smallest value of $n_l$ and $N-n_{l_0} \ll N$. This means that the system is always in the skin phase}.
Since bulk localization is prevented, we expect the OBC and PBC energy spectra to remain disjoint, and no localization should be observed under PBC.\\
{\em (ii) Continuous random disorder.}  In this case the degeneracy of energy spectrum is (almost surely) fully removed and the energy spectrum $E=V_l$ ($l=1,2,3,....,N$) fills the domain $\Omega$ in complex plane, with a probability distribution given by $f(V)$. The eigenvector corresponding to the eigenenergy $E=V_l$ reads
\begin{equation}
\psi_n^{(V_l)}= \left\{
\begin{array}{ll}
0 & 0 \leq n <l \\
1 & n=l \\
\prod_{m=l+1}^n \frac{J}{E-V_m} & l<n  \leq N
\end{array}
\right.
\end{equation}
Note that the eigenstates with $l$ close to $N$ are localized at the right edge of the lattice, i.e. they can be regarded as skin states. However, for small values of $l$, i.e $l < \sim N/2$, this is not necessarily true.   Indeed, if the term
\begin{equation}
I_n \equiv \prod_{m=l+1}^n \frac{J}{E-V_m}
\end{equation}
is asymptotically a decreasing function of $n$ for large $n$, the eigenvector turns out to be mostly localized {\it in the bulk}, while only when $I_n$ secularly grows with $n$ the eigenvector is a skin state, mostly localized at the right edge. In the thermodynamic limit, we can estimate $I_n$ for large $n$ as
\begin{equation}
I_n \sim \exp \{ L(E)n \}
\end{equation}
where 
\begin{eqnarray}
L (E) & = &   \lim_{n \rightarrow \infty} \frac{1}{n} \sum_{m=l+1}^{n} \log  \left( \frac{J}{E-V_m} \right) \nonumber \\
& = &  \int dV f(V) \log \left( \frac{J}{E-V }\right).
\end{eqnarray}
Therefore, when ${\rm Re}\{ L(E) \}>0$, the eigenvector is condensed at the right edge of the lattice, i.e. it is a skin state. On the other hand, when ${\rm Re}\{ L(E) \}<0$ the 
state is localized in the bulk. This means that, if for any energy ${\rm Re}\{ L(E) \}>0$, the system is in the skin phase. Correspondingly, OBC and PBC energy spectra are expected to be disjoint. Conversely, when  ${\rm Re}\{ L(E) \}<0$ for any eigenenergy $E$, in the thermodynamic limit almost any  eigenstate of $H_{OBC}$ is localized {\em in the bulk} of the lattice, i.e. addensation of the eigenstates at the right edge is counteracted and prevented by localization in the bulk  induced by disorder. Therefore, the condition 
\begin{equation}
{\rm Re} \{ L(E) \}= \int dV f(V) \log \left| \frac{J}{E-V }\right| <0
\end{equation}
for all energies $E=V_l$ of the spectrum indicates that the system is in the bulk localized phase. Since in this regime the energy spectrum should become insensitive to the boundary conditions, we expect the PBC energy spectrum to collapse to the OBC energy spectrum. Note that, from the expression of $L(E)$ given by Eq.(10), we expect ${\rm Re}(L)>0$ for weak disorder (skin phase), and ${\rm Re}(L)<0$ for strong disorder (bulk localized phase). In the intermediate region (moderate disorder), we expect that   ${\rm Re}(L)<0$ for some eigenenergies $E$ and  ${\rm Re}(L)>0$ for other eigenenergies. In this case we have coexistence of skin states and bulk localized states for $H_{OBC}$, i.e. the system is in the mixed phase. 

\subsection{Energy spectrum under periodic boundary conditions}
Under the PBC defined by Eq.(2), the matrix $H_{PBC}$ is not anymore upper triangular and so the energy spectrum $E$ is not defined by the elements of the main diagonal. Rather generally, under PBC the system can be found in three different phases, which find a one-to-one correspondence with the three phases of $H_{OBC}$ discussed in Sec.II.A: the {\em bulk localized phase}, the {\em mobility edge phase}, and the fully {\em delocalized phase}.\\
 The {\em bulk localized phase} occurs when almost all of the OBC eigenvectors are localized in the bulk of the lattice, i.e. whenever the condition (11) is satisfied for almost any energy $E$ of the OBC spectrum. Clearly, owing to localization far from the edges of the lattice for almost any eigenvector, in the bulk localized phase the PBC and OBC energy spectra and corresponding eigenvectors do coincide.\\
  The {\em mobility edge phase} corresponds to the coexistence of localized and extended states of $H_{PBC}$. In this phase some (but not all) eigenvectors of the OBC energy spectrum are localized in the bulk. The OBC energies with bulk-localized eigenvectors also belong to the PBC energy spectrum with the same eigenvectors. The other energies $E$ of the PBC spectrum correspond to extended states. The mobility edge phase of $H_{PBC}$ corresponds to the mixed phase of $H_{OBC}$.\\
  The fully {\em delocalized phase} corresponds to the case where all eigenvectors of the OBC spectrum are skin modes, so as the PBC spectrum is fully disjoint from OBC spectrum and formed by extended states solely. The delocalized phase of $H_{PBC}$ corresponds to the skin phase of $H_{OBC}$.\\
  \\
  The energies $E$ of the PBC spectrum, corresponding to extended states, can be formally found by solving an implicit equation, as shown in Appendix A. 
 Like in the previous subsection, it is worth distinguishing two cases.\\
{\em (i) Discrete random disorder.} Let us assume that the on-site potential $V_n$ takes a set of discrete values $W_1$, $W_2$, ..., $W_M$ (the domain $\Omega$ is a set of $M$ points in complex plane) with probabilities $p_1$, $p_2$, .... $p_M$ ($\sum_n p_n=1$). In this case the system is always in the delocalized phase since all OBC eigenvectors are edge states. Thus the energy spectrum of $H_{PBC}$ is outside the domain $\Omega$ and, in the thermodynamic limit, is obtained from the solutions of the transcendental equation
\begin{equation}
\prod_{n=1}^{M}(E-W_n)^{p_n}= J \exp \left(-i q- i \frac{\Phi}{N} \right)
\end{equation}
with $q=q_l=2 \pi l/N$, $l=1,2,...,N$. As discussed in Sec.III below, as the disorder strength is increased  the PBC energy spectrum undergoes spectral deformations from a single circle of radius $J$ into $M$ circles of small radius, centered at the complex energy points $W_1$, $W_2$,..., $W_M$,  corresponding to a sequence of topological phase transitions (Figs.2 and 3). However, for arbitrarily large disorder the small-radius circles never collapse to the OBC energy spectrum, i.e. bulk localization is prevented and $H_{PBC}$ is always in the delocalized phase.\\
{\em (i) Continuous random disorder.} Let us assume that the on-site potential $V_n$ is a continuous random variable with a probability density distribution $f(V)$ in complex plane $V$, with $f(V)=0$ outside a finite domain $\Omega$ of the complex energy plane. In this case the system can be found in either one of the three phases mentioned above, i.e. bulk localized phase, mobility edge phase or delocalized phase.
The portion of the PBC energy spectrum, inside the domain $\Omega$ and corresponding to bulk localized eigenvectors, is the set of energies $E$ such that Eq.(11) is satisfied.  
The portion of the PBC energy spectrum corresponding to extended states is the set of energies $E$ in complex plane satisfying the equation (see Appendix A)
\[
\int dV f(V) \log \left( \frac{ E-V}{J} \right)=-i \left(q+\frac{\Phi}{N} \right)
\]
i.e. 
\begin{equation}
 L(E)=i \left(q+\frac{\Phi}{N} \right).
\end{equation}
with $q=q_l= 2 \pi l/N$ and $l=1,2,...,N$. In the thermodynamic limit, the variable $q$ can be regraded a continuous variable that varies in the $(0, 2 \pi)$ range, and Eq.(13) implicitly defines rather generally one or more sets of closed loops $E=E(q)$ in complex energy plane. A noticeable example, corresponding to a distribution $F(V)$ with radial symmetry, will be presented in Sec.IV.

\subsection{Non-Hermitian winding number}
For any given point gap $E_B$, Ref.\cite{U1} introduced a topological winding number $w_{E_B}$ for the energy spectrum of a disordered lattice under PBC, given by the formula
\begin{equation}
w_{E_B}=\frac{1}{2 \pi i} \int_{0}^{2 \pi} d \Phi \frac{\partial}{\partial \Phi}  \log \det \left\{H_{PBC}(\Phi)-E_B \right\}.
\end{equation}
Basically, $w_{E_B}$ counts the number of times the complex
spectral trajectory encircles the point gap $E_B$ when the flux $\Phi$ is increased from $0$ to $2 \pi$. Clearly, in the bulk localized phase the winding number vanishes, for any point gap energy $E_B$, since the energy spectrum becomes independent of boundary conditions and thus on the flux $\Phi$. As shown in Appendix B, in the delocalized phase the winding number $w_{E_B}$ can be written in terms of the PBC energy spectral curves $E=E(q)$, defined by Eq.(12) or Eq.(13) with $\Phi=0$, as
\begin{equation}
w_{E_B}=\frac{1}{2 \pi i} \int_{0}^{2 \pi} dq \frac{\partial}{\partial q}  \log \left\{E(q)-E_B \right\}.
\end{equation}
 Hence, $w_{E_B}$ counts the number of times the complex eigenenergies $E(q)$ of extended states under PBC encircles the base point $E_B$. Equation(15) also provides the winding number in the mobility edge phase, where $E(q)$ describes the branch of PBC energies of extended states.

\section{Topological phase transitions  without localization for a discrete random disorder}
An interesting result that follows from he general analysis on Sec.II is that, in the case of a discrete distribution of disorder, the PBC energy spectrum undergoes a series of topological phase transitions as the disorder strength is increased, however the system always remains in the fully delocalized phase. This singular behavior was earlier noticed en passant for the special case of binary (Bernoulli) disorder in Ref.\cite{U1}, however we show here that such a result is general to any discrete random disorder. Let us assume that the on-site potential $V_n$ can assume only a finite set of values $W_1$, $W_2$,..., $W_M$ with probabilities $p_1$, $p_2$,..., $p_M$. As shown in Sec.II.A, the PBC energy spectrum is described by the branches $E(q)$ that are the solutions to the transcendental equation (12). Since in the continuous $q$ limit the spectrum is invariant under a change of the flux $\Phi$, we may set $\Phi=0$ in Eq.(12). To show that the system undergoes topological phase transitions as the disorder strength is increased, let us consider two limiting cases. The first one is the vanishing (or weak) disorder limit, i.e. $W_m \ll J$. In this case the PBC energy spectrum consists of only a single closed loop, which is slightly deformed from the circumference of radius $J$. The other limiting case corresponds to strong disorder with $|W_m-W_l| \gg J$ for any $m \neq l$. In this case, an asymptotic analysis of Eq.(12) clearly shows that the PBC energy spectrum is composed by $M$ circles in complex energy plane, centered at the energies $W_1$, $W_2$,..., $W_M$, with small radii $R_1$, $R_2$,..., $R_M$ given by

 \begin{figure}[htbp]
  \includegraphics[width=86mm]{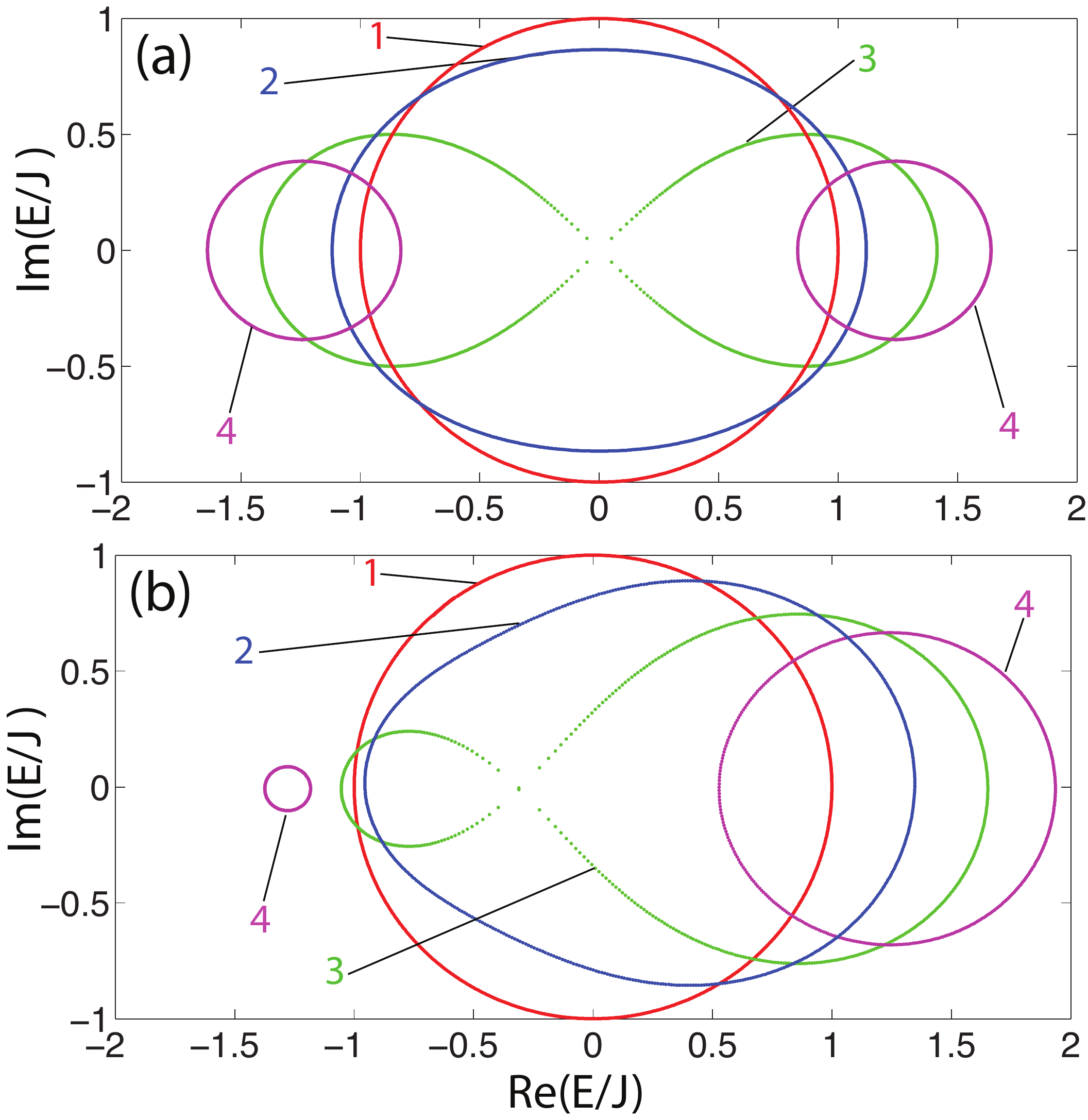}\\
  \caption{(color online) PBC energy spectra in the random dimer model \stef{and in the thermodynamic limit $N \rightarrow \infty$, as predicted by Eq.(16)}. The on-site potential $V_n$ can take only the two values $-W$ and $W$ with probabilities $p$, and $1-p$. In (a) $p=1/2$, whereas in (b)  $p=1/3$. The various curves, labelled by 1,2,3 and 4, refer to increasing values of disorder $W/J$. In (a): $W/J=0$ (curve 1), $W/J=0.5$ (curve 2),  $W/J=W_c/J=1$ (critical point, curve 3), and $W/J=1.3$ (curve 4).  In (b): $W/J=0$ (curve 1), $W/J=0.6$ (curve 2),  $W/J=W_c/J=0.921$ (critical point, curve 3), and $W/J=1.3$ (curve 4).}
\end{figure}
 \begin{figure}[htbp]
  \includegraphics[width=86mm]{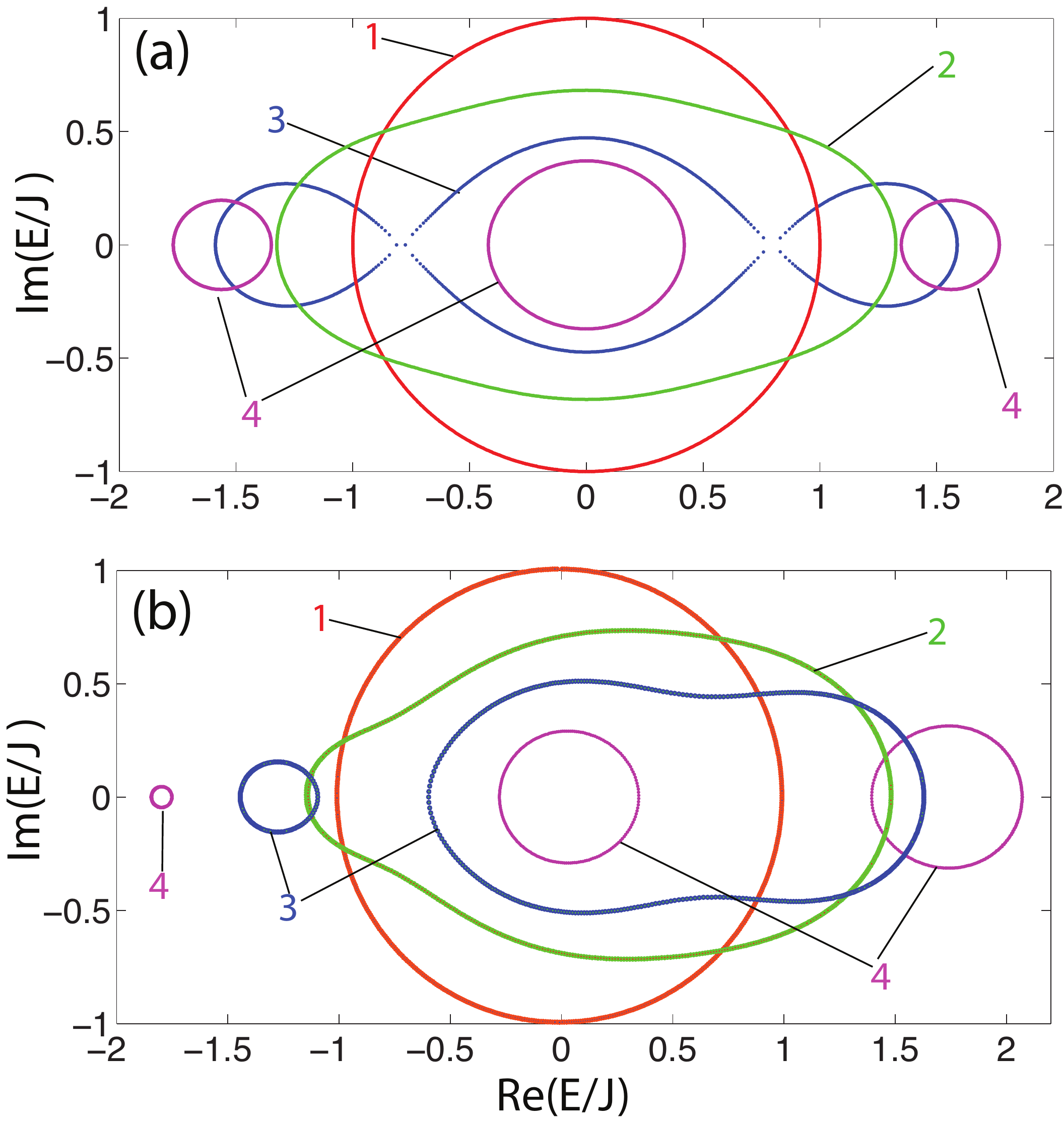}\\
  \caption{(color online) PBC energy spectra in the random trimer model. The on-site potential $V_n$ can take only the three values $-W$, $0$ and $W$ with probabilities $p_1$, $p_2$ and $p_3=1-p_1-p_2$. In (a) $p_1=p_2=p_3=1/3$, whereas in (b)  $p_1=1/4$, $p_2=1/3$ and $p_3=5/12$. The various curves, labelled by 1,2,3 and 4, refer to increasing values of disorder $W/J$. In (a): $W/J=0$ (curve 1), $W/J=1$ (curve 2),  $W/J=W_c/J=\sqrt[3]{13/5}=1.3751$ (critical point, curve 3), and $W/J=1.6$ (curve 4).  In (b): $W/J=0$ (curve 1), $W/J=1$ (curve 2),  $W/J=1.3$ (curve 3), and $W/J=1.8$ (curve 4). The energy spectra \stef{are computed in the thermodynamic limit $N \rightarrow \infty$ from the transcendental equation Eq.(12).}}
\end{figure}\[
R_m= \left[ \frac{J} {\prod_{n=1 \; , n \neq m}^{M} |W_n-W_m|^{p_n}}  \right]^{1/p_m}
\]
($m=1,2,...,M$).
 This means that, as the disorder strength is increased above zero, the original circular loop should break into multiplple loops, indicating the existence of topological phase transitions.\\
To illustrate the results, let us consider in details the PBC spectral deformations in two examples, the random dimer rmodel (Bernoulli disorder)  and the random trimer model.\\
In the random dimer model, the on-site potential $V_n$ can take only the two values $-W$ and $W$ with the probabilities $p$ and $(1-p)$, respectively. The transcendental equation of the PBC energy spectrum reads
\begin{equation}
(E+W)^p(E-W)^{1-p}= J \exp(-iq).
\end{equation}
In the simplest case of equal probabilities $p=1/2$, Eq.(16) can be readily solved yielding
\begin{equation}
E= \pm \sqrt{W^2+J^2 \exp(2iq)}.
\end{equation}
Equation (17) described a single loop for $W<W_c$, and two non-crossing loops for $W>W_c$, where $W_c=J$ is a critical disorder strength corresponding to the loop splitting [Fig.2(a)]. The splitting of the loop is associated to a change of the winding number $w_{E_0}$ [Eq.(15)] with respect to the base energy point $E_B=0$, with $w_{E_0}=1$ for $W<J$ and $w_{E_0}=0$ for $W>J$. Note that such a topological phase transition,  earlier observed in numerical simulations in Ref.\cite{U1}, is not associated to a transition to the bulk phase neither to a mobility edge phase. For $p \neq 1/2$, a similar scenario is observed, but the critical value $W_c$ of the topological phase transition is given by
\begin{equation}
W_c=\frac{J}{2 p^p (1-p)^{(1-p)}}
\end{equation}
while the spectral deformation is not anymore symmetric [Fig.2(b)].\\ 
In the random trimer model, $V_n$ can take only the three values 0, $W$ and $-W$ with probabilities $p_1$, $p_2$ and $p_3$. For $p_1=p_2=p_3=1/3$, the transcendental equation (12) can be reduced to the cubic equation
\begin{equation}
E(E^2-W^2)=J^3 \exp(-3iq)
\end{equation} 
whose solutions are
\begin{eqnarray}
E(q) & = & \sqrt[3]{\frac{J^3 \exp(-3iq)}{2}+ \sqrt{\frac{J^6 \exp(-6iq)}{4}-\frac{W^6}{27}}} \nonumber \\
& + & \sqrt[3]{\frac{J^3 \exp(-3iq)}{2}- \sqrt{\frac{J^6 \exp(-6iq)}{4}-\frac{W^6}{27}}} \;\;\;\;
\end{eqnarray}
In this case it can be shown that a topological phase transition occurs at the critical value
\begin{equation}
W_c= J \sqrt[3]{\frac{13}{5}}. 
\end{equation}
i.e. the PBC energy spectrum consists of a single loop for $W<W_c$, and three non-crossing loops for $W>W_c$ [Fig.3(a)]. If the probabilities $p_1$, $p_2$ and $p_3$ are not equal, as the disorder strength $W$ is increased  one first observes the splitting of the PBC spectrum into two loops, and a further splitting into three non-crossing loops at higher $W$ [Fig.3(b)]. 

\section{Continuous random disorder with radial symmetry: spectral phase transitions and resilience}
In this section we consider the case of continuous random disorder, in which the probability density distribution $f(V)$ of the complex on-site potential $V=R \exp(i \varphi)$ displays a radial symmetry, i.e. it is a function of $R=|V|$ solely. This means that there is equiprobability for the phase $\varphi$ of $V$ to assume any value in the range $(0, 2 \pi)$, while the probability that $R=|V|$ takes a value in the range $(R,R+dR)$ is given by $2 \pi R g(R)dR$, where the distribution $g(R)$ satisfies the normalization conditon
\begin{figure}[htbp]
  \includegraphics[width=86mm]{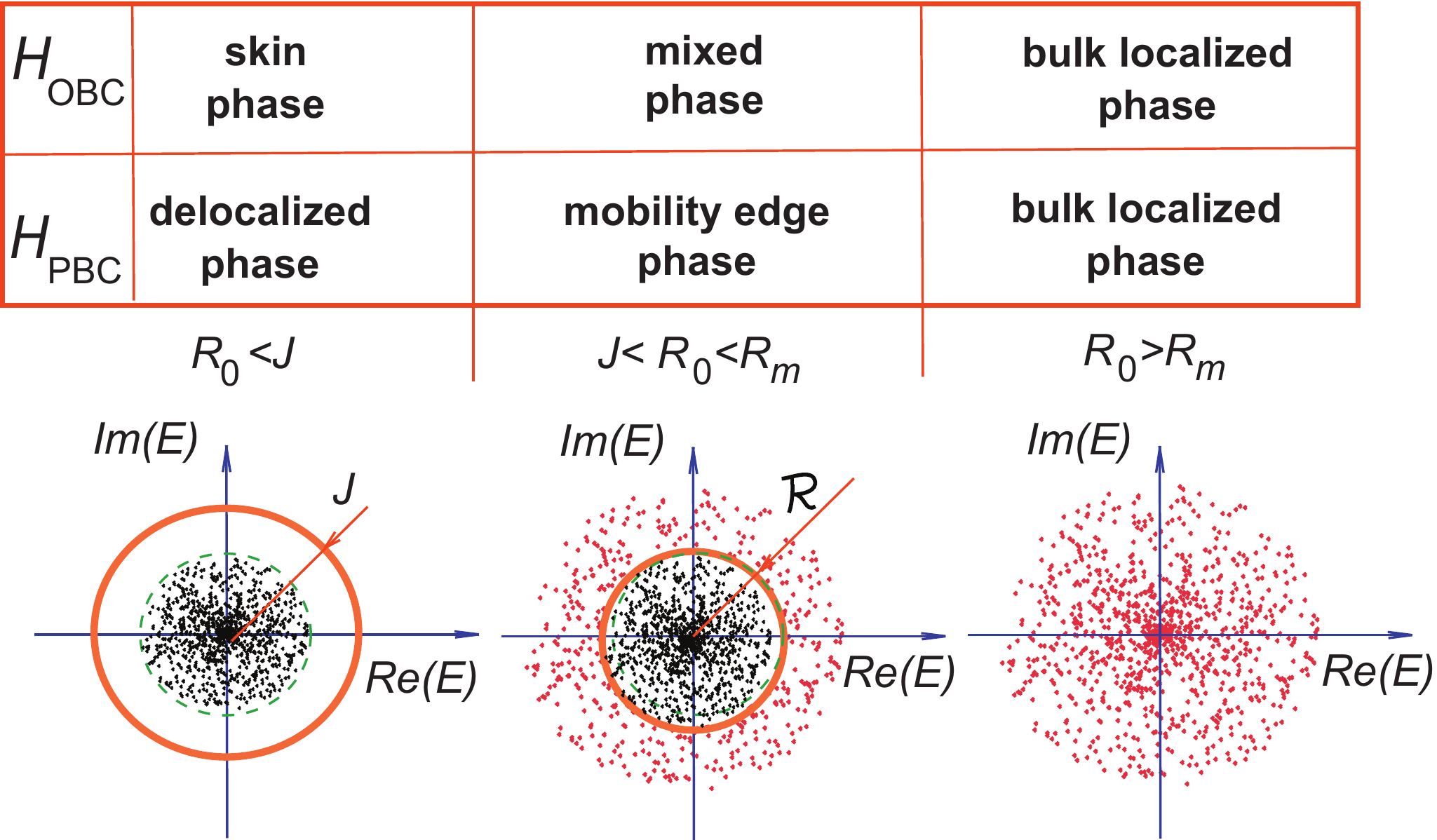}\\
  \caption{(color online) Energy spectra, under OBC and PBC, for a continuous random disorder with radial symmetry, illustrating the existence of three different phases for $H_{OBC}$ and $H_{PBC}$ depending on the strength of disorder $R_0$.
  Dark points correspond to eigenenergies of $H_{OBC}$ solely with eigenmodes squeezed on the right edge of the lattice (skin states), light red points correspond to eigenenergies of both $H_{OBC}$ and $H_{PBC}$ with bulk localized eigenstates, whereas the red circles of radius $J$ and $\mathcal{R}$ in the left and central panels  correspond to eigenenergies of $H_{PBC}$ solely with extended states. The critical disorder strength $R_m$, above which the system enters in the bulk localized phase, is implicitly defined by Eq.(23) given in the text.}
\end{figure}
\begin{equation}
\int_0^ \infty R g(R) dR= \frac{1}{2 \pi}.
\end{equation}
\begin{figure*}[htbp]
  \includegraphics[width=176mm]{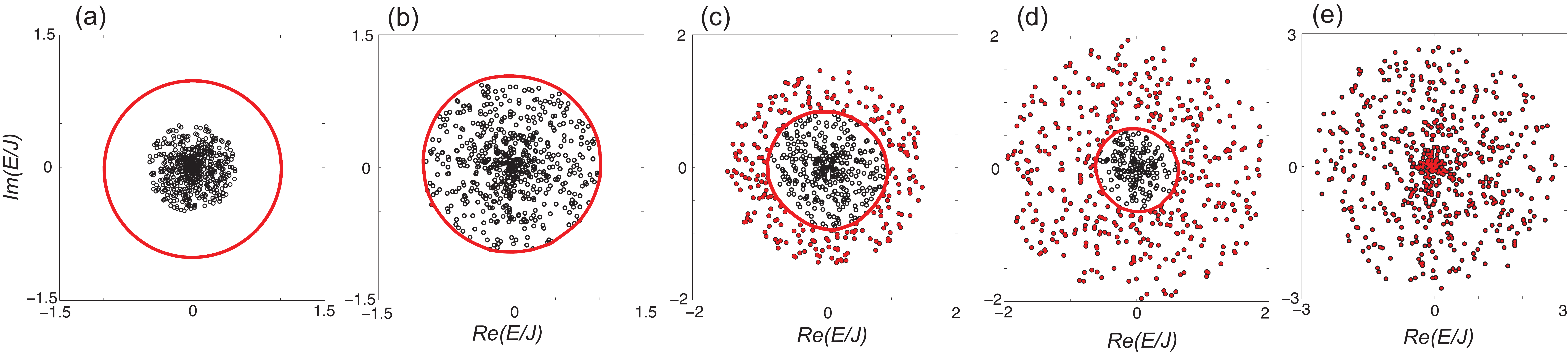}\\
  \caption{(color online) Numerically-computed eigenvalues (energy spectra) of $H_{OBC}$ (open black circles) and $H_{PBC}$  (bold red circles) in a lattice comprising $N=600$ sites for a single realization of on-site potential disorder with radial symmetry. The radial probability distribution of  disorder $g(R)$ is given by Eq.(24) of the main text. In (a) $R_0/J=0.5$, in (b) $R_0/J=1$, in (c)  $R_0/J=1.5$, in (d)  $R_0/J=2$ and in (e)  $R_0/J=2.8$. The critical disorder strength $R_m$, above which the system enters into the bulk localized phase, is $R_m=e J \simeq 2.7813 J$. Note that in (a) $H_{PBC}$ is in the delocalized phase and $H_{OBC}$ in the skin phase, the eigenvalues of $H_{PBC}$ lie on the circumference of radius $J$, with OBC and PBC energy spectra fully disjoint. In (c) and (d) $H_{PBC}$ is in the mobility edge phase and $H_{OBC}$ in the mixed phase, with eigenenergies of $H_{PBC}$ partly distributed along a circle of radius $\mathcal{R}<J$, corresponding to extended states, and partly coincident with eigenenergies of $H_{OBC}$ (the points in the outer of the circle of radius $\mathcal{R}$), corresponding to bulk localized eigenstates. The eigenvectors of $H_{OBC}$ in the interior of the circle of radius $\mathcal{R} $ are skin states.  In (b) the system is at the critical point separating delocalized and nobility edges phases for $H_{PBC}$ (or likewise skin and mixed phases for $H_{OBC}$). In (e) the system is in the bulk localized phase, with equal PBC/OBC energy spectra and all eigenfunctions localized in the bulk.}
\end{figure*}
\begin{figure}[htbp]
  \includegraphics[width=84mm]{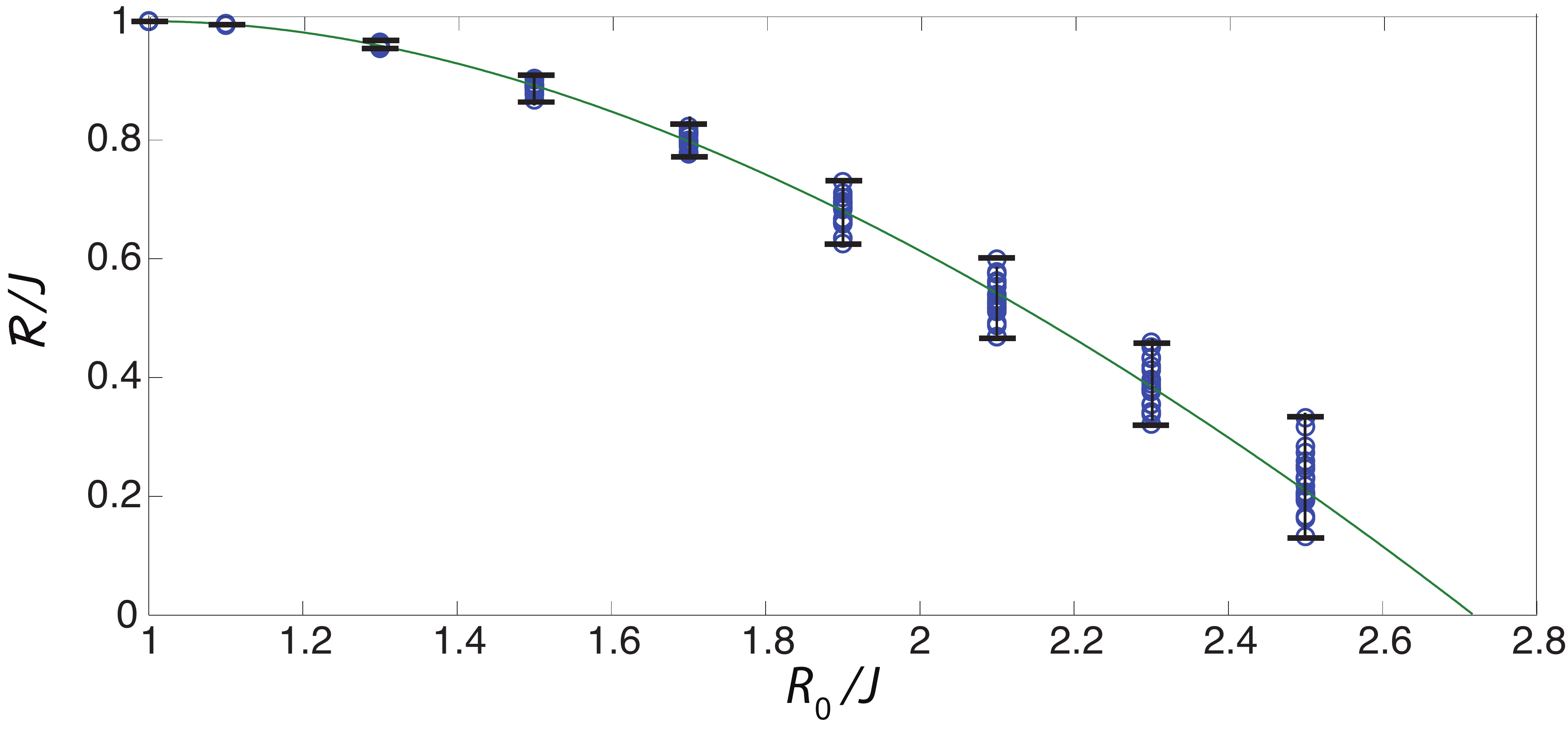}\\
  \caption{(color online) Behavior of the radius $\mathcal{R}$ versus disorder strength $R_0$ for the on-site potential disorder with radial distribution given by Eq.(24) in the mobility edge phase. \stef{Vertical circles refer to numerical simulations for 20 different realizations of disorder in a lattice comprising $N=600$ sites under PBC}, whereas  
  the solid curve is the theoretical prediction of Eq.(26).}
\end{figure}

We also assume that $g(R)$ vanishes for $R \geq R_0$, so that the support $\Omega$ of the disorder is the interior of the circle in complex energy plane of radius $R_0$ and centered at $V=0$. $R_0$ provides a measure of the disorder strength.
As shown in Sec. II.A, the energy spectrum under OBC reproduces the distribution of the on-site potential disorder, while the  PBC energy spectrum undergoes two phase transitions, as stated by following main theorem which is proven in Appendix C and illustrated in Fig.4:\\
(i) For $R_0<J$ (weak disorder regime), $H_{OBC}$ is in the skin phase while $H_{PBC}$ is in the delocalized phase. The PBC energy spectrum lies on the circumference of radius $J$, centered at the zero energy $E=0$, i.e. the disorder does not change the PBC energy spectrum from the disorder-free lattice.\\
(ii) For $J<R_0<R_m$ (intermediate disorder regime), $H_{OBC}$ is in the mixed phase while $H_{PBC}$ is in the mobility edge phase, with coexistence of bulk localized and extended states of $H_{PBC}$, and bulk localized states and skin edge states for $H_{OBC}$. The PBC eigenenergies corresponding to extended states lie on a circumference of radius $\mathcal{R}=\mathcal{R}(R_0)$ smaller than $J$, that shrinks to zero as $R_0 \rightarrow R_m^-$, where $R_m>J$ is defined by the relation
\begin{equation}
2 \pi \int_0^{R_m} dR \; R g(R) \log R = \log J.
\end{equation}
The PBC eigenergies corresponding to bulk localized states do coincide with the OBC eigenenergies $E=V_n$ with $|E|> \mathcal{R}$. The radius $\mathcal{R}=\mathcal{R}(R_0)$ is defined by Eq.(C10) given in Appendix C.\\
(iii) For $R_0>R_m$ (strong disorder regime) $H_{OBC}$ and $H_{PBC}$ are in the bulk localized phase, i.e. all eigenstates are localized in the bulk and OBC-PBC energy spectra do coincide.\\
\\
A main consequence of such theorem, which follows from property (i), is the rather unexpected resilience of the PBC energy spectrum against disorder for $R_0<J$. A further consequence of the theorem is that the winding number $w_{E_B}$ with respect to the point gap $E_B=0$ changes from $w_{E_B}=1$ in the delocalized and mobility edges phases, to $w_{E_B}=0$ in the bulk localized phase.  More generally, for an arbitrary base energy $E_B$ in the interior of the circle $|E|=J$ the change of $w_{E_B}$ from 1 to 0, as $R_0$ is increased, indicates the absence of extended states of $H_{PBC}$ with energies larger, in modulus, than $|E_B|$. In other words, a change of $w_{E_B}$ corresponds to the crossing of boundary separating skin states (with energies $|E|> \mathcal{R}$) and bulk localized states (with energies $|E|<\mathcal{R}$) of $H_{OBC}$. \stef{Finally, it is worth mentioning that the critical value $R_m$ of radial disorder, above which $H_{PBC}$ enters into the bulk localized phase and given by Eq.(23), is obtained by equating the "geometric" mean of the modulus of site-potential energy $R=|V_n|$ with the  "geometric" mean of the hopping amplitude $J$ (which is not fluctuating here). Such a criterion of phase transition bears an interesting analogy with the critical point condition in the random transverse-field Ising spin model \cite{h27,h28,h29,h30}, originally derived by Shankar and Murthy \cite{h27} and later on studied by Fisher using a real-space renormalization-group  method \cite{h28,h29}. In the random transverse-field Ising spin chain, the critical point of para-to-ferromagnetic quantum phase
transition near the zero-temperature  is obtained when the  "geometric" mean of the couplings between sites equals the geometric  mean of the random transverse fields acting on single sites \cite{h29,h30}, which is indeed formally analogous to Eq.(23). Such an analogy could pave the way toward the discovery of interesting connections between non-Hermitian disordered systems and other known integrable models with disorder displaying quantum phase transitions.}\\
To illustrate such results, let us consider as an example the following radial distribution of disorder
\begin{equation}
g(R)= \left\{
\begin{array}{lll}
 \frac{1}{2 \pi R R_0} & {\rm for} & R<R_0 \\
 0 & {\rm for} & R>R_0
\end{array}
\right.
\end{equation}
From Eq.(23), the critical value $R_m$ of disorder amplitude, above which the system enters into the bulk localized phase, can be readily calculated and reads
\begin{equation}
R_m= e J.
\end{equation}
In the mobility edge phase of $H_{PBC}$, i.e. for $J<R_0<R_m$, the radius $\mathcal{R}=\mathcal{R}(R_0)$ of the circumference, that contains the PBC eigenenergies of extended states, is readily calculated from Eq.(C10) given in the Appendix C and reads
\begin{equation}
\mathcal{R}= R_0 -R_0 \log \left( \frac{R_0}{J}\right).
\end{equation}  
Note that $\mathcal{R}$ is a decreasing function of $R_0$, with $\mathcal{R}=J$ at $R_0=J$ (critical point separating delocalized and mobility edge phases) and $\mathcal{R}=0$ at $R_0=R_m= e J$ (critical point separating mobility edge and bulk localized phases). \stef{Note that, for a lattice of finite size $N$, the fractional number $N_{ext}$ of extended states, whose eigenenergies lie on the circle of radius $\mathcal{R}$, is readily calculated as $N_{ext} \simeq N \times 2 \pi \int_0^{\mathcal{R}} dR \;R g(R)=N \mathcal{R}/{R}_0=N- N \log (\mathcal{R}_0/J)$, with $N_{ext} \rightarrow 0$ as $R_0 \rightarrow R_m= e J$.}\\
We checked the predictions of the theoretical analysis by direct numerical computation of eigenenergies and corresponding eigenfunctions for the matrices $H_{OBC}$ and $H_{PBC}$ for a lattice comprising $N=600$ sites, with on-site potential disorder with radial symmetry described by the radial distribution (24). Numerical results of OBC and PBC energy spectra, for increasing values of disorder radius $R_0$, are shown in Fig.5 for a single realization of disorder.  The results are in excellent agreement with the analytical model, in particular they show the resilience of the PBC energy spectrum against disorder in the extended phase $R_0<J$. In Fig.6 we also show the numerically-computed behavior of the radius $\mathcal{R}$ of eigenenergies of extended states of $H_{PBC}$ in the mobility edge phase, for a few values of disorder strength $R_0$ in the range $(J, eJ)$. The radius $\mathcal{R}$ \stef{ is numerically computed from the mean modulus $|E|$ of the energies for the $N_{ext}$ extended states in a lattice comprising $N=600$ sites, and for 20 different realizations of disorder. The $N_{ext}$ eigenenergies  of the PBC spectrum corresponding to extended states form a closed loop that approximately fits the circumference of radius $\mathcal{R}$ predicted by Eq.(26). Both $N_{ext}$ and $\mathcal{R}$ decrease as $R_0$ is increased above $R_0=J$, until to vanish at the critical point $R_0=e J$. The loop turns out to be more sensitive to the realization of disorder as $N_{ext}$ (and thus $\mathcal{R}$) decreases. Such a result explains why the error bars in Fig.6 become wider as the disorder strength is increased from $R_0=J$ to $R_0=2.5 J$}. A comparison of the numerical results with the curve defined by Eq.(26) indicates a very good agreement between theoretical predictions and numerical simulations.

\section{Conclusion}
In this work we presented analytical results on spectral phase transitions and deformations in the non-Hermitian Hatano-Nelson model with unidirectional hopping on the lattice. Despite its simplicity, this model unveils  
interesting effects arising from the interplay between the skin effect and on-site potential disorder. For discrete disorder, the system under PBC is always in the delocalized phase while the system under OBC is always in the skin phase, regardless of the strength of disorder. Moreover, the PBC energy spectrum undergoes a sequence of topological transitions from a single closed loop to a set of non-crossing loops as the disorder strength is increased. For continuous disorder with rotational invariance in complex plane, we predicted rather generally three distinct spectral phases for both OBC and PBC energy spectra. In particular, in one of these phases the disorder does not change the PBC energy spectrum as compared to the disorder-free lattice, revealing a strong resilience of the PBC spectrum against disorder. 

\acknowledgments
The author acknowledges the Spanish State Research
Agency through the Severo Ochoa and María de Maeztu Program for Centers and Units of Excellence in R\&D (Grant No.
FQ 522 MDM-2017-0711).

\appendix
\section{Energy of extended states under periodic boundary conditions}
Let us consider a lattice with PBC and an energy $E$ outside the domain $\Omega$, i.e. distinct than any $V_n$. A formal solution to the difference equation (1) with $J_L=0$ and $J_R=J$ reads
\begin{equation}
\psi_n= \left( \prod_{m=1}^n \frac{J}{E-V_m} \right) \psi_0
\end{equation}
so that the PBC boundary conditions $\psi_{n+N}= \psi_{n} \exp(i \Phi)$ are satisfied provided that
\begin{equation}
\left( \prod_{m=1}^N \frac{J}{E-V_m} \right)= \exp(i \Phi).
\end{equation}
Note that the above equation is an algebraic equation of order $N$ in the energy $E$, which also readily follows from the determinantal equation ${\rm det}(E-H_{PBC})=0$. Thus there are $N$ roots.
If the OBC energy spectrum admits of an eigenenergy $E=V_n$ with corresponding eigenstate localized in the bulk, i.e. ${\rm Re} \{ L(E) \}<0$, it is clear that in the large $N$ limit also Eq.(A2) should admit a solution $E$ that converges toward $V_n$ as $N$ goes to infinity, and the corresponding eigenvector under PBC is also a localized state. Indicating by $N_1 \leq N$ the eigenenergies of $H_{OBC}$ with bulk localized states, i.e. with ${\rm Re}(L)<0$, we are thus left with $N_2=N-N_1$ eigenvectors of $H_{PBC}$ which are extended states. To determine the energies of such extended states in the large $N$ limit, let us take the log of both sides of Eq.(A2) and multiply both sides of the equation so obtained by $(1/N)$. The $N_2$ energies $E$, disjoined from any $V_m$, necessarily have to satisfy the equation
\begin{equation}
\frac{1}{N} \sum_{m=1}^N \log \left(  \frac{J}{E-V_m} \right)= i \left( \frac{\Phi}{N}+q \right)
\end{equation}
where $q=q_l=2 \pi l /N$ and $l=1,2,...,N$. In the large $N$ limit, the term on the left hand side of Eq.(A3) can be approximated by the mean value of the random variable $\log [J/(E-V)]$, i.e one obtains
\begin{equation}
\int dV f(V) \log \left(  \frac{J}{E-V} \right)= i \left( \frac{\Phi}{N}+q \right)
\end{equation}
which yields Eq.(13) given in the main manuscript. Note that, for a discrete distribution of disorder, 
\[
f(V)=p_1 \delta(V-W_1)+p_2 \delta(V-W_2)+...+ p_M \delta(V-W_M)
\]
from Eq.(A4) one obtains
\begin{equation}
 \sum_{m=1}^M p_m \log \left(  \frac{J}{E-W_m} \right)= i \left( \frac{\Phi}{N}+q \right)
\end{equation}
which after some algebraic manipulation yields the transcendental equation (12) given in the main text.

\section{Winding number in the delocalized phase}
The winding number $w_{E_B}$ with respect to a point gap $E_B$ is defined by
\begin{equation}
w_{E_B}=\frac{1}{2 \pi i} \int_{0}^{2 \pi} d \Phi \frac{\partial}{\partial \Phi}  \log \det \left\{H_{PBC}(\Phi)-E_B \right\}.
\end{equation}
Indicating by $E_l=E_l( \Phi)$ the $N$ eigenenergies of $H_{PBC} (\Phi)$, one has
\begin{equation}
w_{E_B}=\frac{1}{2 \pi i} \sum_{l=1}^{N}  \int_{0}^{2 \pi} d \Phi \frac{\partial}{\partial \Phi}  \log \left\{ E_l(\Phi) -E_B \right\}.
\end{equation}
Let us assume that the system, under PBC, is in the delocalized phase, i.e. $E_l (\Phi)$ is distinct than any value $V_n$. Then $E=E_l (\Phi)$ is a solution to Eq.(A3). Clearly, in the large $N$ limit as the flux $\Phi$ varies from $0$ to $ 2 \pi$, the right hand side of Eq.(A3) changes by the infinitesimal amount $ 2 \pi /N$, i.e. the same amount of quantization of $q$. Correspondingly, $E_l (\Phi)$ varies continuously by  a small quantity, and $E_l(\Phi=2 \pi)=E_{l+1}(\Phi=0)$. Therefore Eq.(B3)  can be cast in the form
\begin{equation}
w_{E_B}=\frac{1}{2 \pi i} \sum_{l=1}^{N}   \log \frac{E_{l+1}(\Phi=0)-E_B}{E_{l}(\Phi=0)-E_B}.
\end{equation}
Let us indicate by $E=E(q)$ the solution to Eq.(A3) in the continuous $q$ limit and for $\Phi=0$. Since $E_l(\Phi=0)=E(q_l)$, one has
\begin{eqnarray}
w_{E_B} & = & \frac{1}{2 \pi i} \sum_{l=1}^{N}   \log \frac{E(q_{l+1})-E_B}{E(q_{l})-E_B} \nonumber \\
& = & \frac{1}{2 \pi i}  \int_0^{2 \pi} dq \frac{\partial}{\partial q} \log \{ E(q)-E_B\}
\end{eqnarray}
which is Eq.(15) given in the main manuscript. We note that, if $H_{PBC}$ is in the mobility edge phase, the winding number can be again obtained using Eq.(15) provided that the integral is extended over the loops of the PBC energy spectrum corresponding to extended states solely. 

\section{Energy spectrum for continuous random disorder with radial symmetry}
In this Appendix we prove the main theorem stated in Sec.IV. To this aim, let us assume a radial distribution of the random on-site potential $V=R \exp(i \varphi)$ described by the probability density $g(R)$, with the normalization condition
\begin{equation}
2 \pi \int_0^ \infty dR \; R g(R)=1.
\end{equation}
For a given realization of the disorder, the OBC energy spectrum is given by $E=V_n$. The corresponding eigenfunction, given by Eq.(7) in the main manuscript, is localized in the bulk, and thus it also belongs to the PBC energy spectrum, provided that ${\rm Re}\{ L(E) \}<0$, where $L(E)$ is given by Eq.(10). On the other hand, for ${\rm Re}\{ L(E) \}>0$ the eigenfunction is squeezed toward the right edge, i.e. its is a skin mode, and thus $E$ does not belong to the PBC energy spectrum.

Using polar coordinates, one has
\begin{equation}
L(E)=-\int_0^{\infty} dR \; R g(R) \int_0^{2 \pi} d \varphi  \log \left( \frac{E- R \exp(i \varphi)}{J} \right).
\end{equation}
Taking into account that for any complex number $z$ the following identity holds
\begin{equation}
\int_0^{2 \pi} d \varphi \log \{ 1-z \exp(i \varphi) \}=
\left\{
\begin{array}{lll}
0 & {\rm for} &  |z|<1 \\
2 \pi \log |z| & {\rm for} & |z|>1
\end{array}
\right.
\end{equation}
$L(E)$ can be cast in the form
\begin{equation}
L(E)=- \log \left(  \frac{E}{J}\right)- 2 \pi \int_{|E|}^{\infty} dR \; R g(R) \log \left| \frac{R}{E} \right|
\end{equation}
and thus
\begin{eqnarray}
{\rm Re} \{ L(E) \} & = & - \log \left|  \frac{E}{J}\right|- 2 \pi \int_{|E|}^{\infty} dR \; R g(R) \log \left| \frac{R}{E} \right| \nonumber \\
& = & - 2 \pi \int_0^{\infty} dR \; R g(R) \log \left( \frac{R}{J} \right)+  \nonumber \\
& + & 2 \pi \int_0^{|E|} dR \; R g(R) \log \left|  \frac{R}{E} \right|.
\end{eqnarray}
Note that ${\rm Re} \{ L(E) \}$ is a function of $|E|$ solely, i.e. independent of the phase of the energy $E$, and turns out to be a decreasing function of $|E|$.\\
(i) {\em First part of the theorem}. To prove part (i) of the theorem, let us assume that $g(R)$ vanishes for $R \geq R_0$, with $R_0<J$. This case can be referred to as the $^{\prime}$weak$^{\prime}$ disorder regime.
Then we have 
\begin{eqnarray}
{\rm Re} \{ L(E) \}=- \log \left|  \frac{E}{J}\right|- 2 \pi \int_{|E|}^{R_0} dR \; R g(R) \log \left| \frac{R}{E} \right|   \;\;\;\;\;\;\; \\
> - \log \left|  \frac{E}{J} \right|- 2 \pi \log \left( \frac{R_0}{|E|} \right) \int_{|E|}^{R_0} dR \; R g(R)  \nonumber \\
> -   \log \left|  \frac{E}{J} \right|-  \log \left( \frac{R_0}{|E|} \right)=   \log \left(  \frac{J}{R_0} \right) \nonumber
\end{eqnarray}
i.e. 
\[
{\rm Re} \{ L(E) \} > \log \left(  \frac{J}{R_0} \right).
\]
Hence ${\rm Re}\{ L(E) \}>0$ since $J>R_0$. This means that any eigenfunction of the OBC is not localized in the bulk (it is a skin state), and thus any eigenenergy $E=V_n$ of  
$H_{OBC}$ is not in the spectrum of $H_{PBC}$. Hence the energy spectra of $H_{OBC}$ and $H_{PBC}$ are fully disjoint. To determine the PBC energy spectrum we have to determine the energies $E$ that satisfies the condition (13) given in the main manuscript, i.e.
\begin{equation}
L(E)=i(q+ \Phi/N).
\end{equation}
Taking into account of the form of $L(E)$ [Eq.(C4)], it ready follows that Eq.(C7) is satisfied by letting
\begin{equation}
E(q, \Phi)=\mathcal{R} \exp(-iq -i \Phi/N),
\end{equation}
 where the parameter $\mathcal{R}$, independent of wave number $q$ and of magnetic flux $\Phi$, is found as the root of the equation 
 \begin{equation}
 \log \left( \frac{\mathcal{R}}{J}  \right)+ 2 \pi \int_{\mathcal{R}}^{\infty} dR \; R g(R) \log \left( \frac{R}{\mathcal{R}}\right)=0.
 \end{equation}
 Note that Eq.(C9) is equivalent to state ${\rm Re} \{ L( \mathcal{R}) \}=0$. According to Eq.(C8),  as $q$ varies form $0$ to $2 \pi$ the PBC energy spectrum lies on a circumference in the complex energy plane, centered at the zero energy, of radius $\mathcal{R}$.
 Since $g(R)=0$ for $R \geq R_0$ and $R_0<J$, Eq.(C9) is obviously satisfied by letting $\mathcal{R}=J$. This means that in the weak disorder regime, i.e. when the largest radial disorder of on-site potential $R_0$ is smaller than the hopping amplitude $J$, the PBC energy spectrum is not affected at all by disorder [Fig.4(a)].\\
 (ii) {\em Second and third parts of the theorem}.  Let us now assume $g(R)=0$ for $R>R_0$, with $R_0>J$. Following the reasoning  in the previous proof of part (i) above, the PBC energy spectrum corresponding to extended states, i.e. disjoint from OBC eigenenergies, lies on a circumference of radius $\mathcal{R}$ satisfying Eq.(C9), i.e.
   \begin{equation}
 \log \left( \frac{\mathcal{R}}{J}  \right)+ 2 \pi \int_{\mathcal{R}}^{R_0} dR \; R g(R) \log \left( \frac{R}{\mathcal{R}}\right)=0.
 \end{equation}
 which implicitly defines a function $\mathcal{R}=\mathcal{R}(R_0)$ for $R_0 \geq J$. Clearly, since the second integral on the left hand side of Eq.(C10) is positive, one has $\mathcal{R}(R_0)<J$ for $R_0>J$. Moreover, since the second integral on the left hand side of Eq.(C10) secularly grows with $R_0$, as $R_0$ is increased $\mathcal{R}(R_0)$ should tend to zero. In particular, there will be a value $R_0=R_m$ at which $\mathcal{R}(R_m)=0$. This value is readily obtained noticing that, for infinitesimal $\mathcal{R}$ and $R_0=R_m$, Eq.(C10) can be cast in the form
   \begin{equation}
 2 \pi \int_{{0}}^{R_m} dR \; R g(R) \log \left( \frac{R}{J}\right)=0
 \end{equation}
 which is Eq.(23) given in the main manuscript. This means that, as $R_0$ is increased above $J$, the PBC energy spectrum of extended states, i.e. disjoint from the OBC energies, lies on a circumference of radius $\mathcal{R}(R_0)$ which shrinks toward the zero-energy point as $R_0$ approaches $R_m$. Let us first consider the case $J<R_0<R_m$ [part (ii) of the theorem]. Let $E=V_n$ an energy of the OBC spectrum.  For $|E|>\mathcal{R}(R_0)$, from Eq.(C5) it follows that ${\rm Re} \{ L(E)\}<0$, i.e. the corresponding OBC eigenvector is localized in the bulk and thus $E$ belongs to the PBC spectrum as well. On the other hand, for $|E|< \mathcal{R}(R_0)$ one has ${\rm Re} \{ L(E)\}>0$, so that the OBC eigenvector is not bulk localized (it is a skin state) and $E=V_n$ does not belong to the PBC energy spectrum. Therefore, for $J<R<R_m$, the PBC energy spectrum consists of a set of extended states, with energies $E$ belonging to a circle of radius $\mathcal{R}(R_0)$, and a set of bulk localized states, which are the OBC eigenstates with energies $|E|$ in the outer part of the circle of radius $\mathcal{R}(R_0)$. This proves part (ii) of the theorem of Sec.IV, corresponding to the mobility edge phase [Fig.4(b)]. Finally, let us consider the strong disorder regime  $R_0>R_m$ [part (iii) of the theorem]. In this case, the set of extended states, belonging to the PBC energy spectrum, is empty, and all eigenstates of $H_{OBC}$ are localized in the bulk (there are not skin states), indicating that the PBC and OBC energy spectra do coincide. This regime corresponds thus to the bulk localized phase [Fig.4(c)].

\end{document}